\pdfoutput=1 

\documentclass[a4paper,11pt]{article}

\usepackage[utf8]{inputenc} 
\usepackage{comment}
\usepackage{cite}
\usepackage{multirow}
\usepackage{tabularx}
\usepackage{amssymb}
 \usepackage[pdftex]{graphicx}
 
\usepackage[a4paper, total={6in, 8in}]{geometry}

\title{The Faraday room of the CUORE Experiment}

\begin{document}

\author{C.~Bucci$^{a}$, P.~Carniti$^{b,c}$, L.~Cassina$^{b,c}$, C.~Gotti$^{b,c}$\thanks{Corresponding author}, A.~Pelosi$^d$,\\
G.~Pessina$^{b,c}$, M.~Turqueti$^e$, S.~Zimmermann$^e$}
\maketitle

\footnotesize a) INFN, Laboratori Nazionali del Gran Sasso, Assergi, 67010 L’Aquila, Italy

b) INFN, Sezione di Milano Bicocca, 20126 Milano, Italy

c) Dipartimento di Fisica, Universit\`a di Milano-Bicocca, 20126 Milano, Italy

d) INFN, Sezione di Roma, 00185 Roma, Italy

e) Engineering Division, Lawrence Berkeley National Laboratory, Berkeley, CA 94720, USA

E-mail: {claudio.gotti@mib.infn.it}

\abstract{The paper describes the Faraday room that shields the CUORE experiment against electromagnetic fields, from 50~Hz up to high frequency.
Practical contraints led to choose panels made of light shielding materials.
The seams between panels were optimized with simulations to minimize leakage.
Measurements of shielding performance show attenuation of a factor 15 at 50~Hz, and a factor 1000 above 1 KHz up to about 100~MHz.}


\section{Shielding the CUORE Experiment}

The CUORE experiment is a tonne-scale cryogenic detector searching for rare nuclear decays, located underground at Laboratori Nazionali del Gran Sasso (LNGS) in Italy \cite{CUORE1, CUORE2}.
The main purpose of the experiment is the search for the neutrinoless double beta decay of $^{130}$Te, with a target sensitivity to the half-life of the decay in the range $10^{25}-10^{26}$~year \cite{CUOREsens}.
The CUORE cryostat holds 988 TeO$_2$ crystals at a base temperature close to 10~mK.
The crystals are operated as bolometers, or cryogenic calorimeters: due to their small heat capacity at this temperature ($\sim$10$^{-9}$~J/K) nuclear decays cause measurable thermal signals ($\sim$0.1~mK/MeV) in the crystals.
Thermistors glued to the crystals, and biased with currents in the 100~pA range, convert thermal signals into voltage signals.
These are readout with high gain, low noise voltage amplifiers, operated at room temperature at the top of the cryostat.

The bandwidth of interest for thermal signals is DC up to few tens of Hz.
This is also the frequency range where disturbances must be kept under strict control.
Auxiliary devices in the vicinity of the cryostat (vacuum pumps, etc) can generate low frequency electric and magnetic fields, mostly peaked at 50~Hz and harmonics, which can propagate to the detector by conduction through electrical lines, or irradiated through air.
A proper grounding strategy and a differential readout scheme helps in preventing the former.
The latter can be reduced by shielding with a Faraday cage or, for large volumes (as in the case of CUORE), a Faraday room.

Disturbances at higher frequency (kHz up to GHz) are above the bandwidth of the front-end electronics, and therefore cannot be directly coupled into the detector signals.
However, if they find a way into the cryostat, they can dissipate power, warm up the detectors and degrade the sensitivity.
Therefore, although the focus is at low frequency, the Faraday room should also be effective at higher frequencies.

\begin{figure}[tbp]
\centering
\includegraphics[width=0.8\linewidth]{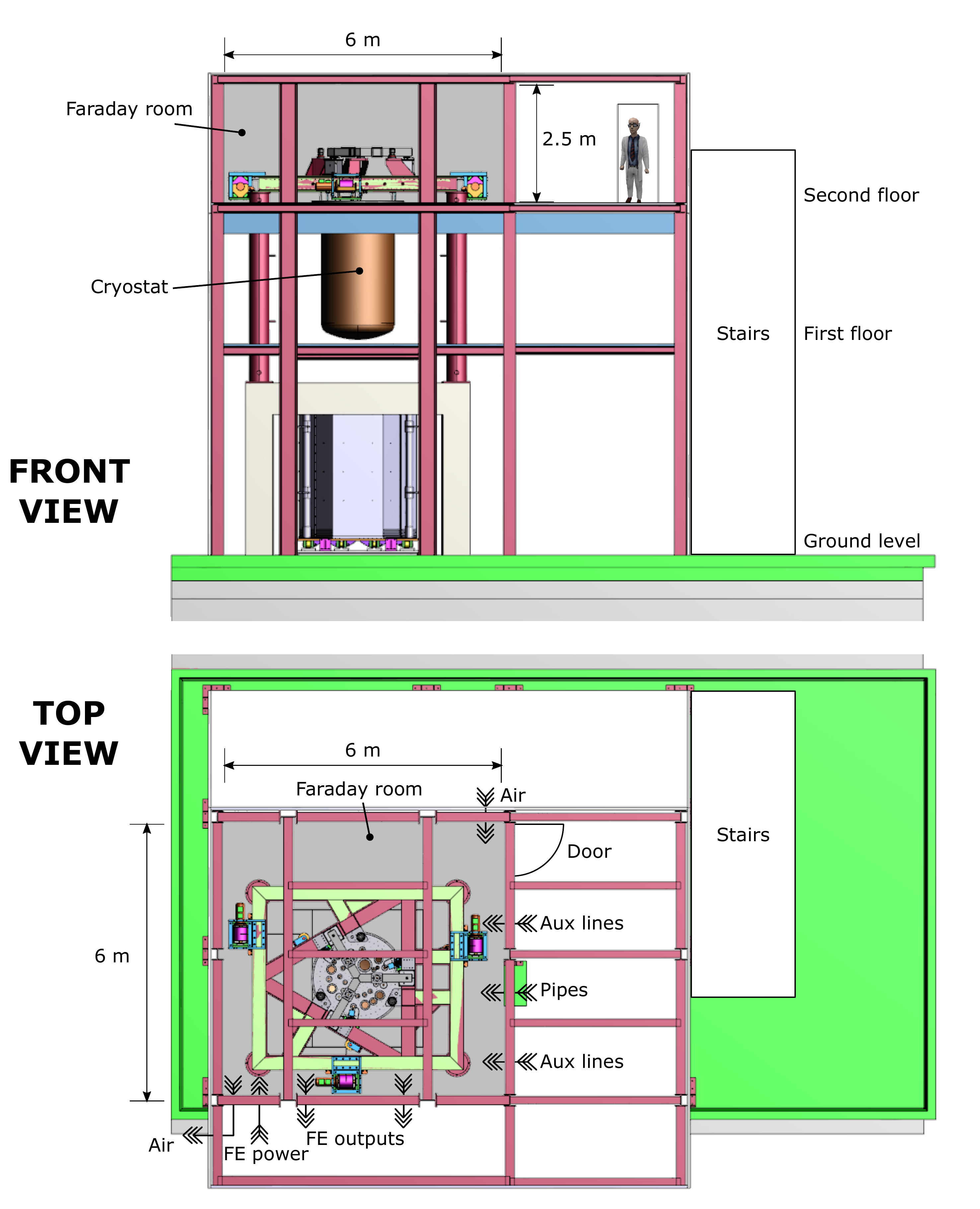}
\caption{The CUORE experimental area, front and top views, with the Faraday room highlighted. Penetrations in the walls of the Faraday room are marked with arrows.}
\label{fig:hut}
\end{figure}

A sketch of the experimental hut at LNGS is shown in figure \ref{fig:hut}.
The Faraday room is located at the top floor of the building, and encloses the area where the top of the cryostat and the front-end electronics are located \cite{frontend, pulser}.
Its floor dimensions are $6\times6$~m$^2$, the internal height is 2.5~m.
The total volume is about 90~m$^3$, and the total surface (walls, floor and ceiling) is close to 130~m$^2$.

Apertures in the walls are needed for pipes and cables (``Aux lines'' and ``Pipes'' in fig \ref{fig:hut}) that connect to the cryostat, for signal cables from the front-end amplifiers (``FE outputs'') and for the power supply lines to the front-end (``FE power'').
The pumping systems are located at the ground floor of the building, as far as possible from the Faraday room, and all the pipes are electrically isolated from the cryostat.
The output signals from the front-end amplifiers cross the wall of the Faraday room before reaching the antialiasing filters and DAQ, located just outside the room, and housed in individually shielded racks.
The DAQ system transmits data through optical fibers to PCs housed in a separate shielded enclosure.
All the power supply lines that enter the room are low noise, regulated, floating DC rails \cite{powersupplies1, powersupplies2}.
The power supply cables that penetrate the room are routed in contact with the floor of the Faraday room and are shielded with a fine mesh.
No AC mains (50~Hz) power supplies are allowed inside the room, and mains routing outside the room was kept as far as possible from the walls.

The floor of the Faraday room is cut to accomodate the cryostat, and is electrically connected to it.
The outer shield of the cryostat is enclosed in a layer of ferromagnetic steel 8~mm thick with relative permeability $\mu_r \simeq 100$, and additional shielding around the cryostat was not deemed necessary (but can be added at a later stage).
The connection to the cryostat brings conflicting requirements: on one side, a good electrical contact without apertures is required for the hermeticity of the electrically shielded volume up to high frequency.
On the other side, since the detector is extremely sensitive to mechanical vibrations, the cryostat must be mechanically isolated from its surroundings, and this includes 
the floor of the Faraday room.
In the present state, a gap is left between the floor of the Faraday room and the cryostat, knowing that this affects the shielding efficacy, especially at high frequency, but allows to prevent any mechanical coupling.
A flexible mesh will be added at a later stage to recover hermeticity, if it will be deemed necessary.

\section{Choice of shielding material}

Detailed presentations of shielding theory and practice can be found in electromagnetic compatibility textbooks \cite{Ott, Paul}.
A brief summary is given here for our purposes.

Far from its source, an electromagnetic wave travels as a plane wave.
Its characteristics depend on the medium where it propagates.
In particular, the ratio between electric and magnetic components E/H is fixed and equal to $\sqrt{\mu/\epsilon}$, where $\epsilon$ and $\mu$ are the dielectric permittivity and magnetic permeability of the medium.
This ``far field'' regime is valid if the source distance is larger than about one wavelength.
In the scale of our system, the former is in the range 1$-$10~m, therefore the ``far field'' regime is valid only in the hundred of MHz or GHz range.
At these frequencies, electromagnetic waves are reflected or absorbed by any continuous conductive surface, but leakage through seams and apertures is most critical.
The shielding efficacy of a Faraday cage in the GHz range is then dominated by the design of seams and apertures (cable feedthroughs, air inlets, joints, etc).

At lower frequencies, at distances of less than one wavelength (``near field'' regime), the characteristics of the field are instead primarily determined by the characteristics of the source.
Fields generated by voltage fluctuations with small currents, e.g. by a rod antenna, have a dominant electric component.
Typical unwanted sources in a laboratory are sparks from gas discharge lamps and the brushes of DC motors.
On the contrary, fields generated by current fluctuations and small voltages, e.g. by a loop antenna, have a dominant magnetic component.
Typical examples are main lines, transformers and motors.

Reflection at the surface and absorption are the physical mechanisms that contribute to shielding.
Their relative importance depends on the nature of the field.
Let us consider the shielding performance of a metal sheet with relative electrical conductivity $\sigma_r$ and relative magnetic permeability $\mu_r$ in the ``near field'' regime:

\begin{itemize}

\item
Reflection occurs at the surface of the shield.
Free charges in the conductor react to the impinging field, creating charge and current distributions that cancel the field at the conductor surface, reflecting it.
For electric fields, this mechanism is effective at all frequencies, down to DC.
The reflection loss is a function of $\sigma / \mu$.
The best performance is provided by non-magnetic materials with high conductivity (e.g. copper, aluminium).
In practice, most metals give a good reflection of electric fields, regardless of thickness.
For magnetic fields, since the efficiency in inducing currents depends on the rate of change of the magnetic field (Faraday's law of induction), reflection is significant only at high frequency, while it is negligible at low frequency.
Reflection therefore does not shield against low frequency magnetic fields.

\item
Absorption occurs in the bulk of the shield.
An electromagnetic wave travelling in a conductive medium induces currents and dissipates energy.
As the field propagates inside the shield by a thickness $t$, it is attenuated by a factor $\exp{\left(-t/\delta\right)}$, where $\delta = 1/\sqrt{\pi f \mu \sigma}$ is the skin depth.
The smallest skin depth at a given frequency is provided by magnetic materials with high magnetic permeability (e.g. iron, mu-metal, metglas), therefore the best shielding performance, although their conductivity is tipically lower than copper or aluminium.
Since reflection is negligible, absorption is the main phenomenon contributing to magnetic shielding at low frequency, even though the skin depth is large.
A shield of significant thickness is required, as the typical skin depth at 50~Hz is of the order of 1~mm in iron and 0.1~mm in mu-metal.

\end{itemize}

To summarize, if mixed electric and magnetic fields are considered, good conductors provide shielding by reflection at high frequency, and for electric field at low frequency, while shielding of magnetic field at low frequency requires absorption by a magnetic material.
To cover a wide frequency range, layers of different materials can be used.
In the case of the CUORE Faraday room, the above guidelines were considered together with specific contraints.
The Faraday room was designed to fit inside the existing CUORE experimental hut, and this resulted in a limit on the total weight of the room.
Moreover, as the access to the experimental area with cranes and similar lifting tools is impractical, the room had to be assembled by hand, setting a further limit on weight.
These constraints forced the choice towards light panels, made of a thin 0.2~mm layer of a high permeability metallic glass alloy named Skudotech, held between two aluminium sheets of 3~mm (outer layer) and 2~mm (inner layer) \footnote{The shielding panels are named Skudal SK3022, manufactured by Selite s.r.l., Assago (MI), Italy.}.
The panels are quoted for a shielding efficacy by absorption of 23 dB (a reduction by a factor 15) for a 50~Hz 10$-$100~$\mu$T field generated by a coil at 20~cm from the panel surface.
The weight of the panels is 15.7 kg/m$^2$.
To obtain the same absorption with a single iron layer, a thickness of $2.5-3$~mm would be needed, with a weight 30\%$-$50\% higher.
The lower weight of the chosen panels with respect to iron allows to build the room using larger panels of 2.5$\times$1.25~m$^2$, weighting 50~kg each.
This reduces the total number of seams, improving shielding performance and simplicity of mounting.

\section{Design of the structure}

\begin{figure}[tbp]
\centering
\includegraphics[width=0.9\linewidth]{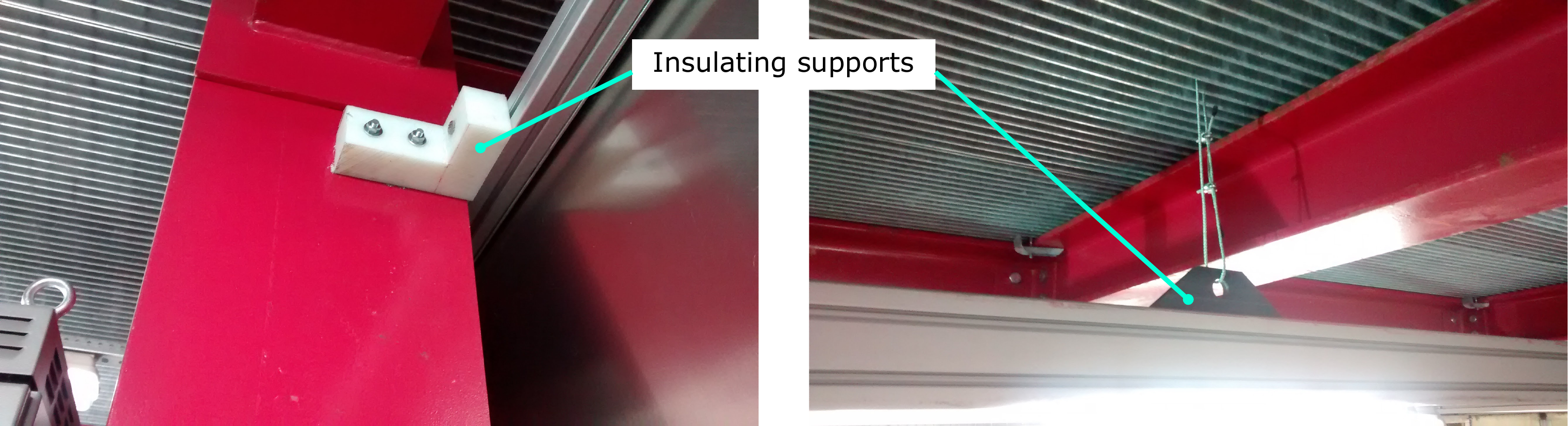} \hspace{1cm}
\caption{The structure that supports the panels is mechanically attached to the beams of the surrounding experimental hut.
Insulating supports for the walls (left) and ceiling (right) prevent electrical contact, to ensure that the Faraday room is properly grounded with the cryostat in only one point.}
\label{fig:supports}
\end{figure}

A support structure made of aluminium profiles was designed to support the panels.
Since the experiment is located in a deep underground laboratory, welding was considered hazardous, and purely mechanical connections were favoured.
The entire structure is electrically isolated from the experimental hut to prevent possible straight currents flowing in the walls of the Faraday room.
The room is electrically connected to the cryostat.
The walls and the ceiling of the Faraday room are attached to the beams of the surrounding experimental hut with insulating supports.
Examples are shown in figure \ref{fig:supports}.

\begin{figure}[tbp]
\centering
\includegraphics[width=0.6\linewidth]{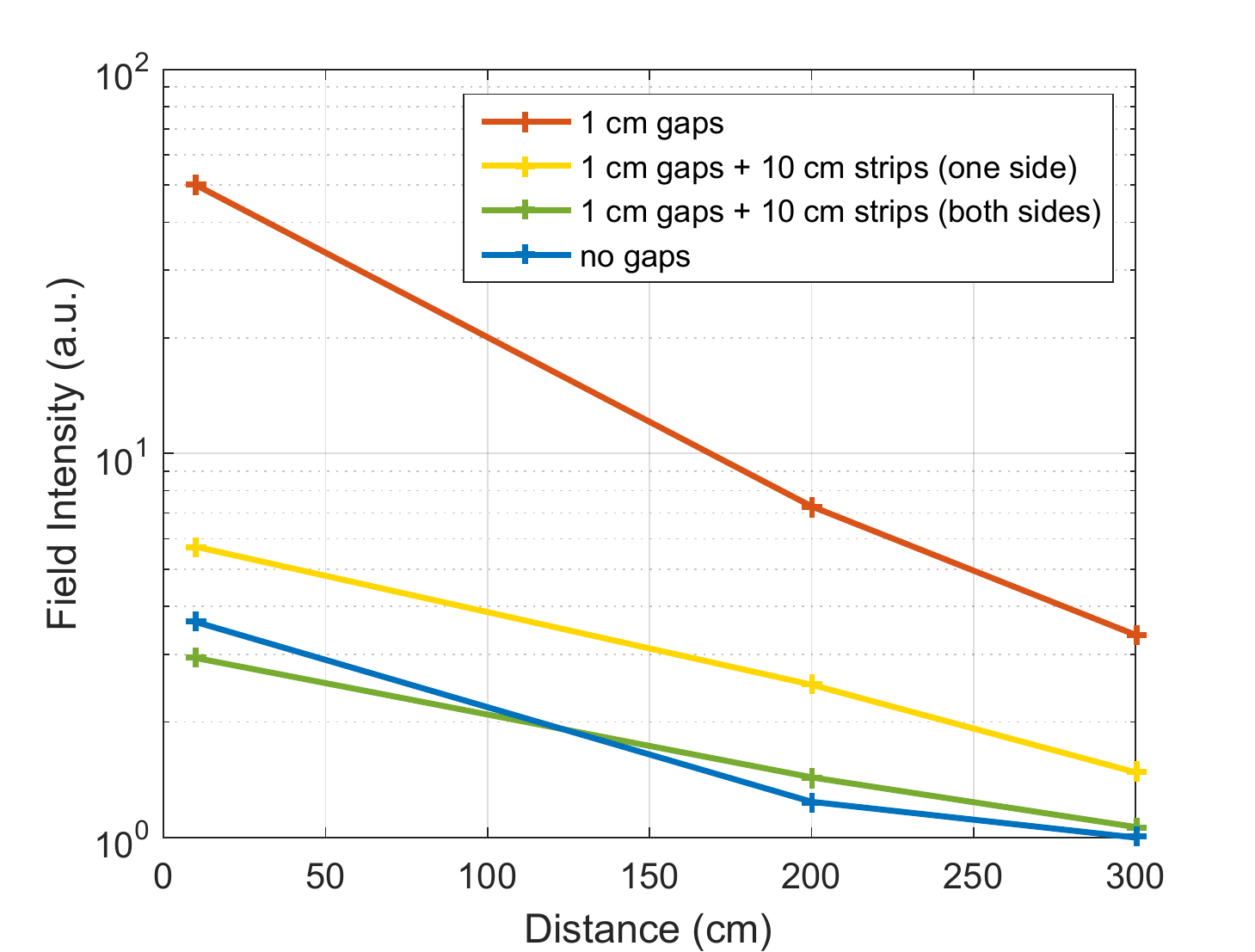}
\caption{Simulation of leakage for the cases described in the text.
The field intensity is normalised to the value obtained with no gaps at 300~cm.}
\label{fig:gapsim}
\end{figure}

The leakage due to gaps between panels was simulated in different conditions, and compared with the ideal case without gaps.
The source of the field in the simulation was a coil generating a 50~Hz magnetic field at 100~cm distance from the wall.
Figure \ref{fig:gapsim} shows the simulated leakage inside the Faraday room as a function of distance from the wall.
The simulation shows that significant leakage is present in the case of 1~cm gaps between panels.
The field intensity close to the wall is almost 20 times larger than in the ideal case with no gaps, and is still three times larger at 300~cm, at the center of the room.
A good shielding performance, within a factor two from the ideal case also closer to the wall, is obtained if the gaps are covered with 10~cm strips of the same material as the panels.
In this case the Skudotech sheets are not in direct contact with each other, and electrical contact between panels is provided only through the aluminium sheets.
An even better performance, almost indistinguishable from the case with no gaps, is obtained if also the other side of the gaps is covered with aluminium strips.
This configuration is actually slightly better than the case with no gaps at 10 cm from the wall, because of the overlap of Skudotech layers, which provide additional absorption.

\begin{figure}[tbp]
\centering
\includegraphics[width=0.9\linewidth]{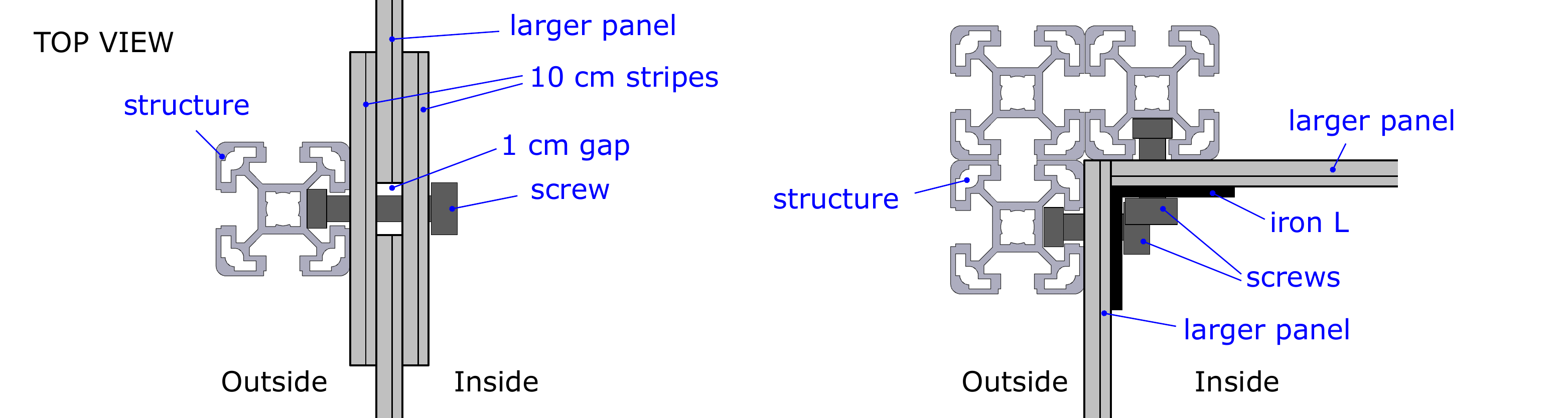}
\caption{Left: The gaps between panels were covered with 10~cm wide strips on both sides to minimise leakage. Right: the corners were covered with L-shaped iron profiles.}
\label{fig:seamdesign}
\end{figure}

Based on these results, the seams between panels were designed as in figure \ref{fig:seamdesign}, left side.
For practical reasons, both strips contain Skudotech, as all the strips were cut from the same material as the main shielding panels.
Most panels and strips were machined at the factory to avoid degrading the magnetic properties of the high permeability shielding material, except a few pieces for the floor, which had to be cut on site to fit properly around the cryostat and its supporting columns.
The edges of the room, where panels meet at 90 degrees, were instead covered with 3~mm L-shaped iron profiles, shown in figure \ref{fig:seamdesign}, right side.

In the long term, the quality of the contacts between aluminium surfaces can deteriorate due to oxidation.
This could impact the shielding performance, especially at high frequency.
To prevent the formation of oxide, all the contact areas were first passed with fine sandpaper, covered with an antioxidant compound\footnote{We used Noalox from Ideal Industries inc., Sycamore, IL, USA.}, and then brushed with a steel wire brush, before screwing them together.
The procedure follows the guidelines for aluminium contacts described in \cite{ContactGuidelines}.

The door of the room is a single 2.5$\times$1.25~m$^2$ shielding panel, framed by the same aluminium profiles that support the room.
The front-end electronics (1000 channels) at the top of the cryostat dissipate almost 1~kW.
When the door is closed, air circulates through two 15 cm diameter holes, marked as ``Air'' in figure \ref{fig:hut}.
The outlet, close to the ceiling, forces air out with a fan.
The fan is mounted on the wall of the hut and connected to the Faraday room with a 1~m duct to prevent injection of 50~Hz interference.
The inlet is at floor level, and allows unforced cool air from outside the hut.
The temperature inside the Faraday room is then only slightly higher than that of the surrounding cavern, and no strong air currents are present inside the Faraday room, which could otherwise induce vibrations to the sensitive area of the experiment.

\section{Tests of shielding efficacy}

The shielding efficacy of the Faraday room was tested by generating a low frequency magnetic field outside the room with a coil, and detecting it inside the room with an identical coil aligned with the first.
Both source and probe were 100~m wires coiled around cylinders of 6.5~cm diameter ($\sim$500~turns).
The two coils were placed across one of the walls, both at 5~cm from the surface.
The probe coil was directly connected to the oscilloscope.
A current in the range 10$-$100~mA was induced in the source coil by a sine wave generator, corresponding to a calculated magnetic field of 30$-$300~$\mu$T.
This field intensity is representative of typical sources of electromagnetic disturbances; for instance, the mains transformer of an instrument located just outside the Faraday room.

\begin{figure}[tbp]
\centering
\includegraphics[width=0.6\linewidth]{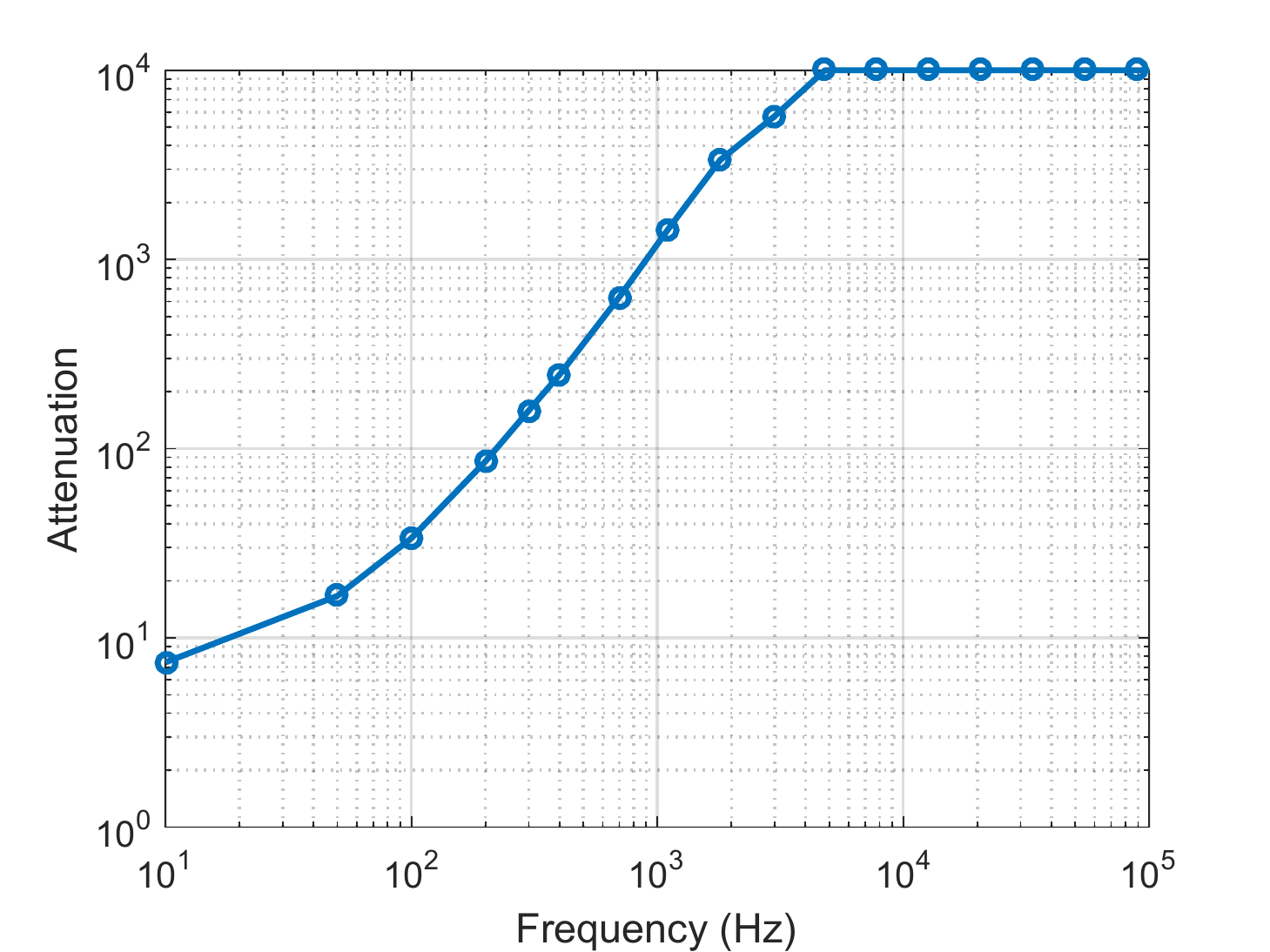}
\caption{Measured attenuation as a function of frequency for magnetic fields generated close to a wall of the Faraday room.}
\label{fig:attenuation}
\end{figure}

The shielding was estimated by comparing the signal detected by the probe in these conditions, with the signal detected with an identical geometry but without the wall between the coils.
The results in the frequency range 10 Hz to 100 KHz are shown in figure \ref{fig:attenuation}.
No dependency on the field intensity was observed, therefore only the curve obtained at $\sim$300~$\mu$T is shown.
Saturation of the magnetic permeability of the 0.2~mm Skudotech layer occurs for field intensities above a few mT, which are not present in the CUORE experimental area.
The shielding is just below a factor 10 (20~dB) at 10~Hz, between 15 and 20 at 50~Hz, and increases reaching a factor 1000 (60~dB) at 1~KHz and 10000 (80~dB) at 5~kHz.
The sensitivity of the measurement is limited to 80~dB by the minimum signal that could be detected at the probe side.
The results at 50~Hz are consistent with what is expected from the shielding panels, and no leakage is observed at the seams.

\begin{figure}[tbp]
\centering
\includegraphics[width=0.9\linewidth]{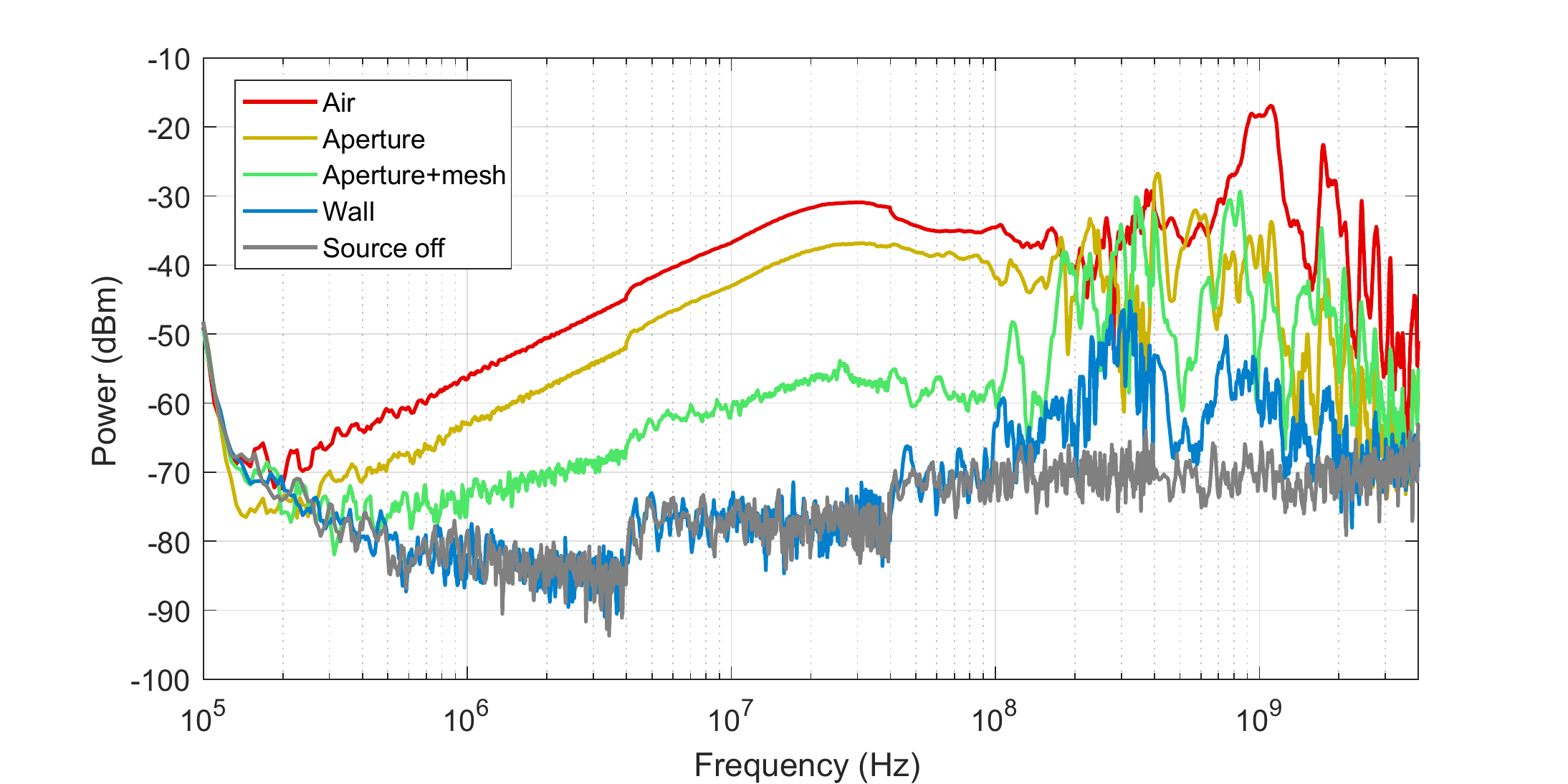}
\caption{Spectra measured by irradiating noise with a loop antenna of 10~cm diameter in different conditions: without any shielding (``Air''), in front of a 3$\times$25~cm$^2$ aperture in the wall of the Faraday room (``Aperture''), with the same aperture covered with a fine copper mesh (``Aperture+mesh''), and through the wall of the Faraday room (``Wall''). The sensitivity of the measurement extends to -80~dBm, as shown in the last curve (``Source off'').}
\label{fig:measRF}
\end{figure}

The shielding efficacy of the room was tested also at higher frequency with a Rohde\&Schwarz FSV4 spectrum analyzer with FSv-B9 tracking generator.
An antenna (a single loop of 10~cm diameter) was connected to the instrument output, generating a RF disturbance in the range 100~KHz $-$ 4~GHz.
An identical antenna aligned with the first at $\sim$10~cm distance was connected to the instrument input and used as a probe.
Figure \ref{fig:measRF} shows the results.
The curve ``Air'' was taken without any shielding, with the noise generator set to the full available power (0~dBm on a 50~$\Omega$ load).
The curve ``Aperture'' was taken in front of a 3$\times$25~cm$^2$ aperture for cables in one of the walls of the Faraday room.
The measured power is only slightly smaller than the previous case, demonstrating considerable leakage.
The situation is improved by covering the aperture with a fine copper mesh (``Aperture+mesh''), although the induced signal is still visible in the entire frequency range.
The wall of the Faraday room (``Wall'') completely blocks the induced field below about 100~MHz, where the measured spectrum is identical to that measured without the noise source (``Source off'').
Leakage can still be observed, but only above 100~MHz.
However, it should be noted that no significant sources of RF noise are present in the underground experimental environment, being shielded by the surrounding rock and ground water, as can be seen by the spectrum ``Source off''.
Similar results were obtained by using rod antennas as both source and probe, in place of loop antennas.

\section{Conclusions}

The design and shielding performance of Faraday room of the CUORE experiment were described.
The room provides shielding by a factor 15$-$20 for 50~Hz magnetic fields, which are the most critical, while satisfying stringent requirements on overall weight and ease of mounting.
The seams between shielding panels were designed to minimize leakage.
High shielding efficacy was measured up to about 100~MHz, although several apertures are needed to operate the experiment, and a trade-off is present between electromagnetic hermeticity and the need to prevent coupling of vibrations to the cryostat.
The shielding provided by the Faraday room is complemented by removing possible sources of disturbance in the vicinity of the room.

\end{document}